\documentclass[twocolumn,showpacs,preprintnumbers,amsmath,amssymb]{revtex4}
\usepackage{amsfonts}
\usepackage{amsmath}
\usepackage{amssymb}
\usepackage{graphicx}
\usepackage{graphics}



\begin{document}

\preprint{submitted to Journal of Applied Crystallography;
published in 2006, \textbf{39}, p.244}

\title{Effects of cryoprotectant concentration and cooling rate on vitrification of aqueous solutions}

\author{Viatcheslav Berejnov}
\email{vb54@cornell.edu}
\affiliation{Physics Department, Cornell
University, Ithaca, NY, 14853}
\author{Naji S. Husseini}
\affiliation{Applied and Engineering Physics Department, Cornell
University, USA}
\author{Osama A. Alsaied}
\affiliation{Weill Cornell Medical College, Doha, Qatar}
\author{Robert E. Thorne}
\affiliation{Physics Department, Cornell University, Ithaca, NY,
14853}


\begin{abstract}
Vitrification of aqueous cryoprotectant mixtures is essential in
cryopreservation of proteins and other biological samples. We
report systematic measurements of critical cryoprotective agent
(CPA) concentrations required for vitrification during plunge
cooling from  T=295 K to T=77 K in liquid nitrogen. Measurements
on fourteen common   CPAs including alcohols (glycerol, methanol,
isopropanol), sugars   (sucrose, xylitol, dextrose, trehalose),
PEGs (ethylene glycol,   PEG 200, PEG 2 000, PEG 20 000), glycols
(DMSO, MPD), and salt (NaCl)   were performed for volumes ranging
over four orders of magnitude from    $\sim$ nL to 20 $\mu$L, and
covering the range of interest in protein crystallography. X-ray
diffraction measurements on aqueous glycerol mixtures confirm that
the polycrystalline-to-vitreous transition occurs within a span of
less than $2\%$ w/v in CPA concentration, and that the form of
polycrystalline ice (hexagonal or cubic) depends on CPA
concentration and cooling rate. For most of the studied
cryoprotectants, the critical concentration decreases strongly
with volume in the range from ~5 $\mu$L to ~0.1 $\mu$L, typically
by a factor of two. By combining measurements of the critical
concentration versus volume with cooling time versus volume, we
obtain the function of greatest intrinsic physical interest: the
critical CPA concentration versus cooling rate during flash
cooling. These results provide a basis for more rational design of
cryoprotective protocols, and should yield insight into the
physics of glass formation in aqueous mixtures.
\end{abstract}

\maketitle

\section{Introduction}
Cryocrystallographic methods play a central role in protein
crystallography (Hope, 1990; Rodgers, 1994; Chayen et al., 1996;
Garman and Schneider, 1997; Garman, 1999; Juers and Matthews,
2004). The amount of data that can be collected from a crystal
before significant degradation in the X-ray beam occurs is
increased and, compared with room temperature data collection in
glass capillaries, crystal handling is in many ways simplified.
These large benefits have accrued even though many aspects of the
cryoprotection and flash cooling processes have been poorly
understood.

Protein crystals are mechanically very fragile. Because they
contain (and can be surrounded by) large amounts of water,
formation and growth of water ice crystals during cooling degrades
protein crystal order and diffraction properties (Garman and
Schneider, 1997). Instead, water must be cooled into an amorphous,
vitreous or glassy state in which the liquid phase's lack of
long-range order is preserved but its molecular motions are frozen
out.

Glass formation during cooling occurs only if crystal nucleation
is somehow suppressed. Because of thermodynamic and kinetic
barriers, formation of crystal nuclei requires a finite time, so
higher cooling rates reduce the probability of nucleation and
favor glass formation. The nucleation probability also depends on
how liquid properties like viscosity vary with temperature, and
thus on the detailed form of the cooling path T(t). Cryoprotective
agents (CPAs) like glycerol, which dissolve in or fully mix with
water, modify molecular diffusion relevant to nucleation. However,
they can also lead to changes in protein structure and, if
introduced via post-growth soaks, can cause osmotic shock and
crystal damage, especially at higher concentrations. The vitreous
ice/polycrystalline ice phase diagram as a function of
cryoprotectant concentration thus provides an important tool for
optimising cryopreservation protocols.

The conventional method used to determine vitreous/crystal phase
diagrams in the temperature/concentration plane is based on
differential scanning calorimetry (DSC) (MacKenzie, 1977). A metal
or glass container holding a solution volume of $\sim10-100$
$\mu$L is plunged into liquid nitrogen to cool the solution into
the vitreous state. Typical cooling rates in DSC experiments are
$<10$ K/s, and the cooling history determines the properties of
the glassy state. Calorimetry measurements are then performed as
the sample is warmed, typically at $\sim1$ K/s. For the slow
cooling rates typical of DSC, homogeneous crystal nucleation
occurs below a critical CPA concentration $C_m$ (Luyet and
Rasmussen, 1968; Rasmussen and Luyet, 1970; Fahy et al., 1984).
Above $C_m$, only heterogeneous nucleation occurs, and this can be
suppressed either by using "clean" samples or by reducing the
sample size. Homogeneous and heterogeneous nucleation can be
distinguished by comparing the behavior of bulk liquid, thin films
(Turnbull and Fisher, 1949), and emulsions (Luyet and Rasmussen,
1968; Rasmussen and Luyet, 1970). $C_m$ thus determines the
minimum CPA concentration required to achieve vitrified ice under
slow-cooling conditions. In protein crystallography, sample
volumes range from $\mu$L to pL - orders of magnitude smaller than
in DSC measurements. The cooling rates therefore are much larger,
so the CPA concentrations needed to achieve vitreous ice should be
smaller.

To date, the kinetic phase diagrams giving the minimum (critical)
CPA concentration required for vitrification during plunge cooling
in common liquid cryogens have not been determined for any CPA in
the volume/cooling rate range relevant to cryocrystallography. In
particular, the development of microfabricated mounts (Thorne et
al., 2003) as a replacement for loops in cryocyrstallography has
allowed very small crystals to be routinely mounted with very
little surrounding liquid, and there is no information on CPA
concentrations required for vitrification in such small volumes.
Recent experiments (Garman and Mitchell, 1996; McFerrin and Snell,
2002; Chinte, et al., 2005) have probed the behavior of larger
volumes obtained by mixing glycerol with Hampton crystallization
screens, with $5\%$ v/v glycerol concentration increments. These
screens contain multiple components such as salts and other CPAs,
and the liquid volumes have not been accurately measured, making
detailed interpretation of results difficult. Critical CPA
concentrations are of interest not just in protein crystallography
but more generally in cryopreservation of biological samples. This
suggests binary aqueous solutions as the starting point for
systematic study.

We have determined the vitreous ice/polycrystalline ice phase
diagrams of CPA concentration versus volume for plunge cooling in
liquid nitrogen. We have examined volumes varying by up to five
orders of magnitude, from $\sim1$ nL (0.1 nL for glycerol) to 20
$\mu$L. The measurements were done for fourteen different
penetrating and nonpenetrating cryoprotectants, including alcohols
(glycerol, methanol, isopropanol), sugars (sucrose, xylitol,
dextrose, trehalose), polyethylene glycols (PEGs) (ethylene
glycol, PEG 200, PEG 2 000, PEG 20 000), glycols (dimethyl
sulfoxide (DMSO), 2-methyl-2,4-pentanediol (MPD)), and salt
(NaCl). Drops of solution were plunge cooled from T=295 K into
liquid nitrogen at T=77 K. The resulting low temperature phase -
vitreous or polycrystalline - was determined from visual
observation of whether the drops remained transparent or became
opaque, supplemented by X-ray diffraction measurements. For the
most common CPAs, the required concentration is a strong function
of volume, and for typical volumes relevant in protein
crystallography is roughly half that required for volumes greater
than 10 $\mu$L. For glycerol, we have also determined the critical
concentration for vitrification versus cooling rate.

\section{Materials and Methods}
\subsection{Preparation of CPA solutions}

All cryoprotective agents were obtained from Sigma-Aldrich (St.
Louis, MO). CPAs were mixed with distilled deionized (DI) water to
produce aliquots in increments of $2\%$ weight per volume (w/v).
For solid CPAs, weighing errors were less than $1\%$. For liquid
CPAs, uncertainties in dilution with DI water in volumetric flasks
were less than $1\%$. To ensure full mixing for solid CPAs near
their solubility limits and for high viscosity liquid CPAs, the
solutions were heated, stirred and sonicated.

\subsection{Drop dispensing, holding and volume estimation}
Four types of sample holder were used to accommodate different
sample volumes: (1) 24 gauge (0.51 mm diameter) copper loops with
1.6-1.7 mm diameters and (2) 36 gauge (0.13 mm diameter) tungsten
loops with $\sim$0.9 mm diameters were used for volumes of 7.5
$\mu$L to 20 $\mu$L and 100 nL to 7.5 $\mu$L, respectively; (3)
CryoLoops (Hampton Research, Laguna Niguel, CA) of 10 $\mu$m nylon
with nominal apertures of 0.1-0.5 mm were used for volumes between
200 and 100 nL; and (4) MicroMounts (Mitegen, Ithaca, NY) made
from 10 $\mu$m polyimide films with apertures of 100 to 500 $\mu$m
were used for volumes smaller than 100 nL. In all cases,
experiments were performed to verify that the sample holder did
not limit cooling rates.

\begin{table}
\caption{\label{tab1}Typical volumes $V$, aspect ratios $R$,
relative volume errors $Er$, dispensers $Dsp$ (a: Pipetman P20, b:
Pipetman P2) and sample holders (1: copper loop, 2: tungsten loop,
and 3: MicroMount, with aperture in $\mu$m, as described in the
text) for the aqueous CPA samples used in these experiments.}
\begin{ruledtabular}
\begin{tabular}{cccccccc}
 $V, \mu L$ & $R$ & Er $\%$ & Dsp & Er $\%$ &  Dsp & Holder\\
\hline
20 & $>0.8$ & 0.5 & a  &  &  & 1 \\
15 & $>0.8$ & 1.5 & a  &  &  & 1 \\
10 & $>0.8$ & 0.5 & a  &  &  & 1 \\
7.5 & $>0.8$ & 1.4 & a &  &  & 1 \\
5.0 & $>0.8$ & 2.4 & a &  &  & 1 \\
2.0 & $>0.8$ & 6.3 & a &  2.2 & b & 2 \\
1.5 & $>0.8$ & 5.6 & a &  5.6 & b & 2 \\
1.0 & $>0.8$ & 16 & a  &  11  & b & 2 \\
0.5 & 0.8 &      &  & 24 & b & 2 \\
0.4 & 1.0 &      &  & 29 & b & 2 \\
0.3 & 0.9 &      &  & 36 & b & 2 \\
0.2 & 0.8 &      &  & 49 & b & 2 \\
0.1\footnotemark[1] & 0.6  &  &  & 77& b & 2 \\
0.050 & 0.4 &    &  & 5  & & 3, 500 $\mu$m\\
0.025 & 0.3 &    &  & 7  & & 3, 400 $\mu$m\\
0.010 & 0.4 &    &  & 8  & & 3, 300 $\mu$m\\
0.005 & 0.4 &    &  & 12 & & 3, 200 $\mu$m \\
0.0002 & 0.3 &   &  & 36 & & 3,\footnotemark[2] 100$\mu$m \\
\end{tabular}
\end{ruledtabular}
\footnotetext[1]{0.5 mm CryoLoop used for X-ray data collection}
\footnotetext[2]{100 $\mu$m MicroMount used for X-ray data
collection}
\end{table}

Larger volume drops were dispensed onto the metal loops using a
pipette (P2 and P20, Gilson, Pipetman, Middleton, WI). As shown in
Figure 1 (a) and (b), the resulting drops were nearly spherical
with very little liquid-metal contact, and had radii slightly
larger than the loop radii.  Small volume drops were prepared by
immersing CryoLoops or MicroMounts directly into CPA solutions.
CryoLoops yielded nearly spherical drops. The smallest drops
prepared using MicroMounts were flattened in the plane of the
circular aperture, with aspect ratios of roughly $0.3-0.4$. To
determine drop volumes captured by CryoLoops and MicroMounts, the
drops were treated as oblate spheroids, and the dimensions of
their axes measured using a microscope. The volume was calculated
as $V=(4/3)\pi x^2z$, where $x$ is the radius in the plane of the
aperture and $z$ is half of the drop's height. The error in the
volume was calculated using standard error propagation based on
the error in the dimensions of the drops. Table I shows typical
measured drop dimensions and calculated volumes obtained for
samples prepared using MicroMounts.

\subsection{Cooling and visual phase inspection}
Flash cooling was performed by immersing samples in liquid
nitrogen (LN2) contained in a glass hemispherical dewar (no.8130,
Pope Scientific Inc, Saukville, WI). In some experiments the
nitrogen was cooled below its boiling point by reducing the
pressure above the LN2/vapor interface. The dewar was sealed with
an o-ring by a transparent, 1 cm thick Plexiglass lid, and
connected to a vacuum pressure station equipped with a pressure
gauge (Barnat). After pumping, boiling in the open dewar was
eliminated. However, no appreciable difference in flash cooling
results for boiling (T=77 K) and cooled LN2 was observed, so most
data was collected at T=77 K.

Drops were plunged by hand from room temperature (T=295 K) into
LN2 within a second after mounting, to prevent any condensation or
evaporation that could change the sample composition and volume.
The plunge depth was $\sim$1-2 cm below the LN2 surface and the
plunge speed was $\sim$10 cm/s. To maximize heat transfer, drops
were agitated at $\sim$1 cm/s for several seconds until all
boiling had ceased. The boiling time for drop plus holder ranged
from a few seconds for the largest drops in copper wire holders to
a time shorter than could be resolved by eye for the smallest
drops in MicroMounts. Initially, boiling involved nearly periodic
evolution of single bubbles 2-3 mm in diameter. These bubbles were
much larger than the drop size and formed a film around the drop
during their growth.  We identify this as the film-boiling regime
(Incropera and DeWitt, 1981; van Stralen and Cole, 1979). Boiling
behavior then showed an abrupt and visually obvious transition to
simultaneous evolution of several small bubbles with sizes much
smaller than the drop size, which we identify as the bubble
nucleation regime. Each region produces characteristic sounds as
the bubbles break the LN2 surface, and the transition between them
can easily be heard. For larger drop volumes ($>0.5$ $\mu$L) we
were able to accurately measure the time between immersion and
this audible transition, which we hereafter refer to as the
boiling time $t_{boil}$. Since the film-boiling regime lasted much
longer than the bubble nucleation regime, $t_{boil}$ gives a good
estimate of the total cooling time. Based on measurements of drops
plunge-cooled in thermocouples (Teng and Moffat, 1998), we
estimate that the boiling time is roughly twice the time to cool
from room temperature to water's glass transition temperature
$T_g\sim$140 K.

The drop opacity/transparency was ascertained by optical
observation through a stereo-microscope (StereoZoom 6, Bausch and
Lomb, Rochester, NY) with fiber optic illumination (Schott -
Fostec). To facilitate observations, the solidified drop was
raised to just below the LN2 surface, and a black paper background
inserted behind it to enhance contrast.

\subsection{X-ray diffraction measurements}
X-ray diffraction was used to confirm that the transparent/opaque
transition corresponded to the glass/crystal transition, and to
characterize the phase of the smallest samples for which
ascertaining transparency was difficult. Samples were flash cooled
in LN2 and then mounted using CryoTongs (Hampton Research) (to
minimize the temperature rise during transfer) in a nitrogen gas
cryo-stream at T=100 K. X-rays were produced by a Rigaku Rotaflex
rotating anode X-ray generator operating at 50 kV and 100 mA, and
detected using a Rigaku R-Axis $IV++$ detector located 200 mm
away. Exposure times were one minute for $1^\circ$ oscillations,
and yielded a d-spacing resolution of less than 2 $\AA$. Raw X-ray
images corresponding to amorphous or polycrystalline states of the
solidified drops were analyzed using HKL2000 (Otwinowski and
Minor, 1997) and Datasqueeze (Datasqueeze Software, University of
Pennsylvania, PA).

Diffraction patterns were acquired for glycerol-water mixtures
with drop volumes of 100 nL (in 0.5 mm CryoLoops) and 200 pL (in
100 $\mu$m MicroMounts.) The first glycerol concentration tested
was deep in the opaque regime ($22\%$ w/v for 200 pL and $16\%$
w/v for 100 nL), and the concentration was then increased in $2\%$
w/v increments through the critical concentration and into the
transparent regime ($30\%$ w/v for 200 pL, and $58\%$ w/v for
100nL). Figure 2 shows typical diffraction image sequences at 100
nL and 200 pL, respectively.

\section{Results}
\subsection{Vitreous/polycrystalline phase diagrams and critical}

Visual observation in a microscope proved to be an efficient and
reliable way to identify the vitreous-to-polycrystalline ice
transition. As shown in Figure 1, the transition from transparent
to opaque drops was dramatic and occurred without an intermediate
"cloudy" phase. For a given CPA and drops of the same volume, the
drop-to-drop variation in CPA concentration at the transition was
less than $4\%$ w/v, and variations presumably reflected
differences in drop shape and/or the details of how it cooled.

For larger drops held in wire loops, as the CPA concentration was
decreased through the transition region, a patch of opaque ice
first formed in the region farthest from the wire while the rest
of the drop remained transparent. This indicates that heat
transfer through the wire was more efficient than directly from
drop to nitrogen. The patch grew with decreasing CPA concentration
until it consumed the entire drop. The critical CPA concentration
was determined as the smallest concentration that did not yield a
repeatable patch of opaque ice. Similar behavior was not observed
for smaller drops mounted in CryoLoops or MicroMounts.

Figure 3 shows the resulting data for the critical CPA
concentration required for vitrification versus drop volume, for
volumes ranging from $\sim$3 nL to $\sim$30 $\mu$L. The 14 CPAs
are grouped into five categories: alcohols (glycerol, methanol,
isopropanol), sugars (sucrose, xylitol, dextrose, trehalose), PEGs
(ethylene glycol, PEG 200, PEG 2 000, PEG 20 000), glycols (DMSO,
MPD), and salt (NaCl). Concentrations at and above the indicated
critical concentrations yielded transparent, vitreous drops, while
concentrations below yielded opaque, polycrystalline drops.

\begin{table*}
\caption{Minimum cryoprotective agent concentrations (in units of
$\%$ w/v) for vitrification of 14 CPAs versus volume. Values in
units of $\%$ v/v can be obtained by dividing $\%$ w/v by the CPA
density $\rho_{CPA}$ $(g cm^{-3})$ at $T\sim22^\circ$C listed in
the bottom row, obtained from the material safety data sheets. For
a given volume, drops flash cooled with CPA concentrations smaller
than those listed yield crystalline ice.  If no concentration is
listed, then the saturation limit was reached before vitrification
could be achieved. The "Ratio" of minimum concentrations at the
largest and smallest volumes is listed at the end of the table.
Gly=glycerol, Meth=methanol, Isop=isopropanol, Sucr=sucrose,
Xyl=xylitol, Dext=dextrose, Treh=trehalose, and EthGly=ethylene
glycol.}
\begin{ruledtabular}
\begin{tabular}{ccccccccccccccc}
$V,\mu$L& Gly & Meth  &Isop & DMSO & MPD & NaCl & Sucr & Xyl &
Dext & Treh & Eth Gly & PEG 200 & PEG 2000 & PEG 20000 \\
\hline 20 & 52 &  30 & 17.5 & 48 & 27.5  & & 58 & 48 & 32 & 74 &
50 & 50 & 48 &  48\\
15 & 52 & 30 & 17.5 &   48  &27.5 & & 58 & 46 & 30 & 72 & 48 & 50&
48 & 48\\
10 & 52 & 30 & 17.5  &  48 & 27.5  && 56 & 44 & 30  & 72 & 48 & 50
& 48 & 48\\
7.5 & 52 &  27.5  &  17.5  &  46  & 27.5  & & &  & 30 & &46 &48
&48 & 48 \\
5.0 & 50 & 27.5 &   17.5 &   46 & 27.5 & & 54 &44  &30  &62 & 44&
48& 46 &48\\
2.0 &48 & 27.5 &   17.5  &  42&  27.5 & & & 42&  28&  56&  40& 46&
46&48\\
1.5 &48&27.5&17.5&40& 25 & &52 & 40& 28& 54& 38& 44 &44
&48\\
1.0 &42& 27.5& 17.5& 38&  25& & 50&  40&  28&  54&  38&  42&  42&
48\\
0.5 & 40 & 27.5 & 15 & 38 & 22.5 &&&&&50 & 36& 40& 42&48\\
0.4 &38  &25  &15 & 38 & 22.5 & &46 & 36 & 26 & 50 & 36 & 40 & 42
&48\\
0.3 &36 & 25 & 15 & 38 & 22.5&&44&&&48 & 36 & 38 & 40 &48\\
0.2 &36&25& 15& 36& 17.5& &44& 36& 24& 48& 30& 38& 40&  48\\
0.1 &32&25 &15 & 28 & 17.5&& 40 & 32 & 22 & 46 & 26 & 38 & 38 & 48\\
0.050 & 28 & 25 & 15 & 28 & 15 &&36 & 32 & 20 & 44 & 26&&&\\
0.025 &  28 & 25 & 12.5 &   28 & 15 & 22 & 36 & 32&  18 & 44 & 26&&& \\
0.010 &  28 & 25 & 12.5 &   28 & 15 & 22 & 36 & 32 & 18 & 44 & 26&&& \\
0.005 & 28  &25  &12.5 &   28&  15 & 18&&&&&&&&\\
0.0002 &  28 &&&&&&&&&&&\\
Ratio &  1.86  &  1.20  &  1.40  &  1.71  &  1.83  &  1.22  & 1.61
&1.5 &1.78 &   1.68  &  1.92  &  1.32  &  1.26  &  1.00 \\
$\rho_{CPA}$ & 1.26  &  .792  &  .785  &  1.1 &.93 &2.16 &   1.59
&1.52 &1.54  &  1.53  &  1.12 &   1.12 &   1.2 & 1.2\\

\end{tabular}
\end{ruledtabular}
\end{table*}

Most of the studied cryoprotectants show three regimes of
behavior: a large volume ($>10\mu$L) region where the critical
concentration is nearly constant; an intermediate volume region
where the critical concentration decreases with decreasing volume;
and a small volume ($<0.1\mu$L) region where the critical
concentration again becomes nearly constant. Table II gives the
critical concentration of each CPA versus sample volume, and the
ratio of the critical concentrations in the large and small volume
regimes, respectively.

For NaCl, the solubility limit was reached before vitrification
could be achieved at larger volumes. NaCl is a particularly poor
cryoprotectant, because it is effective only at small volumes and
even then at concentrations that are inhospitable to most proteins
and cells. For the PEGs, increasing the molecular weight extended
the large-volume region of nearly constant critical concentration
to lower volumes, and PEG 20,000 did not show any volume
dependence between 0.1 $\mu$L to 20 $\mu$L.

\subsection{Characterization of the vitreous ice/polycrystalline ice
transition by X-ray diffraction}

X-ray diffraction was used to verify that the transparent/opaque
optical transition indeed corresponds with the vitreous ice /
polycrystalline ice transition. Vitreous ice (Dowell and Rinfret,
1960) produces a diffuse ring of scattering near 3.7 $\AA$,
corresponding to the mean intermolecular spacing in the vitreous
phase. Polycrystalline hexagonal ice produces a sharp ring at 2.24
$\AA$ (110) and a cluster of three rings ((100), (002), and (101))
with the central ring at 3.7 $\AA$ (Dowell and Rinfret, 1960;
Murray et al., 2005; Kohl et al., 2000). In the following we refer
to the rings at 2.24 and 3.7 $\AA$ as $\alpha$ and $\beta$ rings,
respectively. The (220) planes of cubic ice produce a ring at 2.24
$\AA$, and the (111) planes produce one strong ring (as opposed to
3 rings) at 3.7 $\AA$. The number of rings at the $\beta$ ring
spacing can thus be used to distinguish the crystalline symmetry
of the ice, provided that finite-size broadening determined by the
crystallite sizes is smaller than the hexagonal ring spacing.

In previous studies, X-ray diffraction measurements of the
vitreous - polycrystalline transition used relatively large (0.1-1
$\mu$L) and ill-defined drop volumes (Garman and Mitchell, 1996;
McFerrin and Snell, 2002; Chinte et al., 2005). We examined
smaller volumes of 100 nL and 200 pL, which correspond to the
range of volumes relevant to protein crystallography on
$\sim30-300$ $\mu$m crystals.

Figures 2 and 4 show a subset of a series of diffraction patterns
obtained as the glycerol concentration was increased from zero and
through the optically determined critical concentration. For both
100 nL and 200 pL volumes, the diffraction patterns of pure water
showed scattered spots at the $\alpha$ and $\beta$ d-spacings of
hexagonal ice, indicating the presence of a finite number of
relatively large crystallites. When glycerol is added, these spots
become more numerous and eventually broaden into continuous rings,
indicating an increase in the number of randomly-oriented
hexagonal crystallites and thus a decrease in their size. On
further concentration increase, the three hexagonal $\beta$ rings
are replaced by a single, much more intense ring (inset c),
indicating a phase change from hexagonal to cubic ice (Dowell and
Rinfret, 1960; Murray et al., 2005; Kohl et al., 2000). Above a
critical glycerol concentration, the $\alpha$ ring disappears and
the $\beta$ ring becomes broad and diffuse (insets b and a),
indicating a transition to vitreous ice.

\subsection{Cooling rate measurements}
Figure 5 shows the boiling time $t_{boil}$ versus drop volume V
for a $40\%$ w/v glycerol solution, where $t_{boil}$ was
determined as the time after a drop was submerged that the audible
(and visual) transition from film boiling to nucleate boiling was
observed, as discussed in Section 2.3. For moderate volumes the
data are well described by $t_{boil}\sim V^{1/2}$ (indicated by
the solid line a), and for large volumes by $t_{boil}\sim V^{2/3}$
(indicated by the solid line b). Open circles show previous data
obtained by directly measuring the cooling time $t_c$ of 0.2 and
0.8 $\mu$L drops using a thermocouple (Teng and Moffat, 1998).
They are in good agreement with our data, confirming the validity
of our measurements even at the shortest cooling times we
measured.

\section{Discussion}
\subsection{Behavior versus concentration near the critical concentration}
The simple method of optical observation (McFerrin and Snell,
2002) proved remarkably effective for vitreous/polycrystalline
phase identification of CPA solutions over the
four-order-of-magnitude volume range explored. This method gives
an upper bound on the width of the transition region between
polycrystalline and vitreous phases of $<4\%$ w/v. X-ray
measurements on 100 nL glycerol-containing drops further narrow
the span of this transition region to less than $2\%$ w/v,
corresponding to $\sim6\%$ of the $32\%$ critical concentration.
The number of water molecules per glycerol molecule thus changes
from 11.7 at the resolved upper concentration limit of the
polycrystalline phase to 10.8 at the resolved lower limit of the
vitreous phase, for the cooling rate achieved in the 100 nL drop.

\subsection{Evolution of ice structure with cryoprotectant concentration}
The cubic crystalline form of ice is not observed during slow
cooling of water or water-glycerol solutions. For pure water,
cubic ice is formed by warming vitreous ice above water's glass
transition temperature $T_g$. Usually cubic ice then converts to
hexagonal ice at a higher temperature, so there is a thermotropic
(temperature-dependent) cubic-to-hexagonal transition (Dowell and
Ringfret, 1960; Murray et al., 2005). In our experiments we
observe a robust lyotropic (concentration-dependent)
cubic-to-hexagonal transition, controlled by the glycerol
concentration. For both the 200 pL and 100 nL drops, the cubic
phase is achieved on flash cooling over a wide glycerol
concentration range, suggesting that the cubic ice phase is
kinetically favored during flash cooling and/or that the presence
of glycerol modifies the preferred equilibrium ice structure. A
similar transition sequence from vitreous to cubic to hexagonal
ice during flash cooling has been observed as a function of
sucrose concentration (Lepault et al., 1997).

\subsection{Volume dependence of the critical CPA concentrations}
For most CPAs in Figure 3, the measured variation of the critical
concentration C separating vitreous and polycrystalline phases
with drop volume shows three regimes. At large volumes and thus
slow cooling rates, the critical concentration for all CPAs is
nearly constant. This is consistent with $C\longrightarrow C_m$
(Rasmussen and Luyet, 1970; Fahy et al., 1984), where $C_m$ is the
minimum CPA concentration for which homogeneous crystal nucleation
is completely inhibited and the liquid phase cools directly into
the vitreous phase, regardless of cooling rate.

At intermediate volumes, the critical CPA concentration shows a
sharp decrease with decreasing volume. The cooling rate increases
with decreasing volume, which reduces the time during which
nucleation and growth of crystalline ice can occur. For a given
CPA concentration, the vitreous phase is kinetically favored
beyond a critical cooling rate. Recent measurements (Chinte et
al., 2005) on mixtures of glycerol and crystal screen solutions
placed in three sizes of CryoLoops (of decreasing but ill-defined
volume) show a similar trend of decreasing critical glycerol
concentration with decreasing volume. An important exception to
this general behavior is provided by higher molecular weight PEGs,
which show little or no decrease in concentration with volume. The
large volume/slow cooling rate critical concentration $C_m$ in
mg/mL (~monomers per unit volume) is roughly the same for the
various chain lengths, but the shorter chains are more effective
in suppressing ice formation at smaller volumes/larger cooling
rates. This latter trend is consistent with Garman's observation
for $~\sim0.2-1$ $\mu$L drops that smaller $M_W$ PEGs are more
effective cryoprotectants (Garman and Mitchell, 1996).

However, for small volumes $(<0.1$ $\mu$L) the critical
concentration levels off and saturates, contrary to naive
expectations that smaller volumes should cool faster. This
saturation could result either because $i)$ the cooling rate
saturates, or $ii)$ because a minimum amount of cryoprotectant is
required to prevent crystalline ice formation, regardless of the
cooling rate.

The second of these hypotheses is inconsistent with experiments
showing that pure water can be vitrified if cooling rates are
large enough (drop volumes are small enough) (Angell, 2002). As
will be discussed elsewhere, to test the first of these
hypotheses, CPA-containing drops were sprayed onto the bottom of
small cups made from ultra-thin copper foil, which were then
immersed in liquid nitrogen. Heat transfer from the drop then
occurs primarily via its contact with the copper rather than via
nitrogen vapor or liquid. For volumes of $0.1-10$ $\mu$L, the
critical CPA concentrations measured in this way are a few percent
lower than those in Figure 3, but show the same decrease with
decreasing drop volume. However, unlike for drops directly plunged
in liquid nitrogen, no plateau is observed at small volumes.
Instead, the behavior observed at intermediate volumes continues
to the smallest drops (1 pL) measured. This suggests that a
saturation in cooling rate is responsible for the small-volume
plateau in Figure 3.

\subsection{Critical concentration and critical cooling rate
glass/crystal diagram}

The measurements of the boiling time in Figure 5 can be used to
determine the cooling rate as a function of drop volume. We define
an average cooling rate $q$ as $q=\Delta T/t_c$, where $\Delta
T=(295-77)K=218K$ and $t_c$ is the time to cool to $T=77 K$.
Assuming $t_c\sim t_{boil}$ and noting that $\Delta T$ in our
experiments is constant, Figure 5 corresponds to a plot of the
reciprocal cooling rate $q^{-1}(V)$.

By combining $q(V)$ obtained from Figure 5 with the critical
concentration $C(V)$ from Figure 3, we can eliminate the volume
$V$ and obtain the minimum cryoprotectant concentration to achieve
vitrification as a function of cooling rate, $C(q)$.  This is the
function of greatest intrinsic physical interest: $C(V)$ depends
on the cooling method (gas stream versus liquid plunge) and medium
(nitrogen, propane, ethane or helium), whereas $C(q)$ is nominally
independent of these parameters and directly connects to the
microscopic kinetics responsible for glass formation.  It can be
used to determine the critical concentration for any cooling
method, medium or sample volume.  To obtain this function, we use
the fits $t_{boil}\sim V^{1/2}$ for $0<V<5$ $\mu$L and
$t_{boil}\sim V^{2/3}$ for $5<V<100$ $\mu$L in Figure 5 and $C\sim
C_m(1-e^{-Va})b$ for $C(V)$ in Figure 3. We exclude data from the
small volume saturation region.

This experimentally determined function for glycerol is shown in
Figure 6. The black circles represent our data and the solid line
is a guide to the eye. Below the line the cooling rate and
cryoprotectant concentration are too small to eliminate
homogeneous nucleation of crystal ice, and above this line
vitreous ice is obtained.

Figure 6 also shows previous data for the critical cooling rates
for water-glycerol mixtures reported by Sutton (Sutton, 1991a;
Sutton, 1991b), indicated by the open circles, and by Kresin and
Korber (Kresin and Korber, 1991), indicated by the open box. These
data were obtained by DSC using surfactant-containing emulsions of
glycerol-water drops. Sutton's data in the low-cooling-rate,
high-concentration region show an increase in critical
concentration with decreasing cooling rate, in contrast to the
clear plateau evident in our results. This discrepancy may be due
to the surfactant, which may affect nucleation. On the other hand,
Kresin and Korber's single data point at high cooling rates lies
well above our data, suggesting either that the cooling rates
actually achieved in their experiment were much lower than
reported, or that assumptions in modelling the fraction of
crystalline ice present within the vitreous phase were incorrect.

\subsection{Comparison with previous work}
The basic methods used in the present work - X-ray diffraction
measurements (Garman and Mitchell, 1996; McFerrin and Snell, 2002;
Chinte et al., 2005) and visual observations (McFerrin and Snell,
2002) - have been used in previous studies of the effects of CPAs
on vitrification of cryoprotective solutions used in protein
crystallography. Garman and Mitchell determined glycerol
concentrations required to vitrify 50 crystallization solutions
from Hampton Research in a nitrogen cryostream at $T=100$ K. Drops
were mounted in single size $(\sim1 mm)$ mohair loops, and their
volumes, although not reported, are estimated to be $\sim0.2-0.6$
$\mu$L. McFerrin and Snell extended Garman and Mitchell's
experiments to determine critical concentrations of four CPAs
(glycerol, PEG 400, ethylene glycol and propylene glycol) required
to vitrify 98 different crystallization solutions from Hampton
Research, using 0.7-1 mm CryoLoops and nitrogen cryostream
cooling. The critical concentrations determined for the single
volume examined in both these studies for drops containing a
single cryoprotectant are consistent with the present results,
although the present results provide better resolution ($2\%$ w/v
vs. $6\%$ w/v). Recently, Chinte et al., (2005) used CryoLoops of
sizes of 1, 0.5, and 0.1 mm to again determine critical glycerol
conditions in a T=100 K nitrogen cryostream, for 12 different
crystallization solutions from Hampton Research. Sample volumes
were not measured, but the expected decrease in critical glycerol
concentration with loop size was observed.

\section{Conclusion}

We have described the first quantitative and systematic
measurements of critical cryoprotectant concentrations for plunge
cooling in liquid nitrogen in the volume range relevant to
cryocrystallography and to the cryopreservation of small numbers
of cells. For most of the fourteen CPAs examined, the critical CPA
concentration is constant at large volumes, decreases strongly
with decreasing volume for intermediate volumes, and then has a
plateau at small volumes. We have combined these measurements of
critical concentration versus volume with measurements of cooling
times versus volume to obtain the quantity of greatest intrinsic
physical interest in flash cooling: the minimum CPA concentration
for vitrification versus cooling rate. Our results can be used to
develop more rational approaches to cryopreservation of protein
crystals and other biological samples. They also provide a
starting point for understanding the physics of flash cooling and
vitrification of aqueous cryoprotectant mixtures.

\begin{acknowledgments}
The authors would like to thank Dr. Jan Kmetko, Eugene Kalinin,
and Dr. Angela Toms for their assistance in operating the X-ray
diffractometer and in interpreting the resulting images. This work
was funded by the National Institute of Health (R01 GM65981).
\end{acknowledgments}

\section*{Figure Captions}

Figure 1\\
Flash-cooled glycerol solutions above and below the critical
concentration for vitrification.  a) 500 nL of $42\%$ w/v glycerol
on a tungsten loop; b) 500 nL of $38\%$ w/v glycerol on a tungsten
loop; c) 50 nL of $30\%$ w/v glycerol on a 500 $\mu$m MicroMount;
d) 50 nL of $26\%$ w/v glycerol on a 500 $\mu$m MicroMount. The
critical concentrations for 500 nL and 50 nL are $40\%$ w/v and
$28\%$ w/v , respectively. The bar represents 500 $\mu$m.\\

Figure 2\\
X-ray diffraction images above and below the critical
concentration of glycerol. The images on the left are from a 200
pL volume in a 100 $\mu$m MicroMount. The images on the right are
from a 100 nL volume in a 0.5 mm Hampton CryoLoop. The rings
marked $\alpha$ and $\beta$ are located at 2.24 $\AA$ and 3.70
$\AA$, respectively. The innermost ring spanning 6.5 $\AA$ to 8.0
$\AA$ in the 200 pL images is from the MicroMount, and is visible
because of the very small sample volume and area within the much
larger area of the X-ray beam.\\

Figure 3\\
Linear and semi-logarithmic plots of the critical concentration
for vitreous ice formation for 14 cryoprotectants. Closed squares
represent: PEG200 (a, b), xylitol (c, d), glycerol (e, f), and
DMSO (g, h); open squares represent: ethylene glycol (a, b) and
sucrose (c, d); closed triangles represent: PEG 20,000 (a, b),
trehalose (c, d), isopropanol (e, f), and NaCl (g, h); open
triangles represent: PEG 2000 (a, b), dextrose (c, d), methanol
(e, f), and MPD (g, h).\\

Figure 4\\
Normalized X-ray diffraction intensity versus d-spacing of 100 nL
flash-cooled drops containing glycerol concentrations of a) $58\%$
w/v, b) $32\%$ w/v, c) $30\%$ w/v, and d) $16\%$ w/v. Curve $a$
corresponds to the vitreous (transparent) phase; curve $b$ is
close to vitreous ice-crystalline ice transition; curve $c$
corresponds to cubic ice or a cubic ice - vitreous ice mixture;
and curve $d$ corresponds to hexagonal ice. \\

Figure 5\\
Boiling time versus drop volume for pure water plunged in liquid
nitrogen on linear and (inset) log-log scales. Each point (solid
circle) represents the average of three independent measurements,
with an average residual of $\sim0.16 s$. Open circles correspond
to data from Teng and Moffat (1998) for 0.2 and 0.8 $\mu$L drops
plunge cooled in LN2. The fit $a$ to data at moderate volumes
corresponds to  $t_{boil}\sim V^{1/2}$, and the fit $b$ to data at
large volumes to $t_{boil}\sim V^{2/3}$, respectively. \\

Figure 6\\
Critical concentration versus cooling rate for glycerol on semilog
and (inset) linear scales. Solid circles are from the present
measurements, and open circles and boxes are from DSC measurements
of Sutton (1991) and Kresin and Korber (1991). The solid line is a
guide to the eye.\\

\end{document}